\documentclass{pnastwo}

\usepackage{graphicx}
\usepackage{amssymb,amsfonts,amsmath}

\contributor{}
\url{}
\copyrightyear{}
\issuedate{ }
\volume{}
\issuenumber{}

\begin{document}


\title{Emergence of good conduct, scaling and Zipf laws in human behavioral sequences in an online world}

\author{
Stefan Thurner\thanks{To whom correspondence should be addressed. E-mail:stefan.thurner@meduniwien.ac.at}\affil{1}{Section for Science of Complex Systems, Medical University of Vienna, Spitalgasse 23, 1090 Vienna, Austria}
  		\affil{2}{Santa Fe Institute, 1399 Hyde Park Road, Santa Fe, NM 87501, USA}
        		\affil{3}{IIASA, Schlossplatz, A-2361 Laxenburg, Austria}, 
Michael Szell\affil{1}{}\and 
Roberta Sinatra\affil{4}{Dipartimento di Fisica e Astronomia, Universit{\`a} di Catania and INFN, Via S. Sofia, 64, 95123 Catania, Italy}
			 \affil{5}{Laboratory for Complex Systems, Scuola Superiore di Catania, via San Nullo 5/i, 95123 Catania, Italy}
}

\contributor{}

\maketitle


\begin{article}

\begin{abstract} 
We study behavioral action sequences of players in a massive multiplayer online  game.
In their virtual life players use eight basic actions which allow them to interact with each other.
These actions are communication, trade, establishing or breaking friendships and enmities, attack, and punishment.
We measure the probabilities for these actions conditional on previous taken and received actions and find 
a dramatic increase of negative behavior immediately after receiving
negative actions. Similarly, positive behavior is intensified by receiving positive actions.
We observe a tendency towards anti-persistence in communication sequences. 
Classifying actions as positive (good) and negative (bad) allows us to define binary `world lines' of 
lives of individuals. Positive and negative actions are persistent and occur in clusters, indicated by  large scaling
exponents $\alpha \sim0.87$ of the mean square displacement of the world lines.
For all eight action types we find strong signs for high levels of repetitiveness, especially for negative actions.
We partition behavioral sequences into segments of length $n$ (behavioral `words' and `motifs') and study their statistical
properties.  We find two approximate power laws in the word ranking distribution, one with an exponent of
$\kappa \sim-1$ for the ranks up to 100, and another with a lower exponent for higher ranks.
The Shannon $n$-tuple redundancy yields large values and increases in terms of word length, further  underscoring the non-trivial
statistical properties of behavioral sequences. 
On the collective, societal level the timeseries of particular actions per day can be understood by a simple mean-reverting log-normal model.
\end{abstract}


\dropcap{S}ocieties can be seen as individuals interacting through a {\em multiplex network} (MPN), i.e. a superposition of several social networks defined on 
the same set of nodes (individuals) \cite{wassermann, PNAS}. Different types of networks correspond to different types of social interactions.
For example the communication sub-network of the MPN
is the network whose links correspond to the exchange of information by means of  emails,  telephone calls, or  letters.
Another subnetwork is the trading network, where goods or services are exchanged between individuals, in exchange for other goods,  money, 
or --rarely-- for nothing. 
Each of these interactions usually needs an initial action taken by one of the subjects involved in the exchange,  the {\em sender}, and a target to receive it,  the {\em recipient}.
Actions can (but do not have to) be reciprocated, so that in general the MPN consists of a set of directed and weighted subnetworks. 
The MPN is a highly non-trivial dynamical object. 
The different social networks within the MPN are not independent but strongly influence each other through a {\em network-network interaction}. 
To understand systemic properties of societies it is essential to detect and quantify the organizational principles behind such  mutual influences. 
The MPN is an example of a {\em co-evolving} structure: on one hand the actions of individuals shape and 
define the topological structure of the MPN. On the other hand the topology of the MPN constrains and influences the possible actions which 
take place on the MPN. In general the MPN of a society can not be observed due to immense requirements on synchronized data acquisition. 
Despite these difficulties, the analysis of {\em small-scale} MPNs has a tradition in the social sciences
\cite{wassermann,mcpherson,entwisle,padgett}. Concerning large-scale studies, 
recently there have been significant achievements in understanding a number of massive social networks on a quantitative basis, such as the 
cell phone communication network \cite{onella1,onella2,renaud}, 
features of the world-trade network \cite{hausmann}, email networks \cite{email},  the  network of financial debt \cite{banks} and the 
network of financial flows \cite{flow}. 
The integration of various dynamical networks of an entire society has so-far  been beyond the 
scope of any realistic data  source. However with the  increasing availability of vast amounts of electronic fingerprints people leave
throughout their lifes, this situation  is about to change.  
Online sources are capturing more and more aspects of  life, boosting our understanding of collective human behavior \cite{lazer2009css, lewis2008ttt}.
One particular source where {\em complete} behavioral multiplex data is available on the society level are massive multiplayer online games (MMOGs). 
In MMOGs hundreds of thousands of players meet online in a `virtual life' where their actions can be easily studied \cite{bainbridge2007srp}. 
Players have to gain their living through economic activity and usually are integrated in several types of social networks. In such games communication 
networks, friendship and enmity networks have been studied, initially as separated entities
\cite{grabowski2007ess,SN}. In \cite{PNAS} trading, aggression and punishment networks have been added to the analysis and  first measurements on  mutual 
network-network influences  were reported. 

In this paper we 
do not focus on the full MPN but on the dynamics (actions) taking place on its nodes. 
We report on the nature of sequences of human behavioral actions in a virtual universe of a MMOG. 
There sequential behavioral data  is available on the scale of an entire society, which is in general impossible to obtain.  
The unique nique data of the  online game \verb|Pardus|
 \cite{pardus} allows to unambiguously track all  actions of all players over long time periods. 
We focus on the stream of eight types of  actions  which are translated into an $8$-letter alphabet. 
This {\em code}  of actions of individual  players is then  analyzed by means of standard timeseries 
approaches as have been used, for example, in DNA sequence  analyzes \cite{dna, dna2, dna3}. 

\begin{figure}[t]
    \begin{center}
             \includegraphics{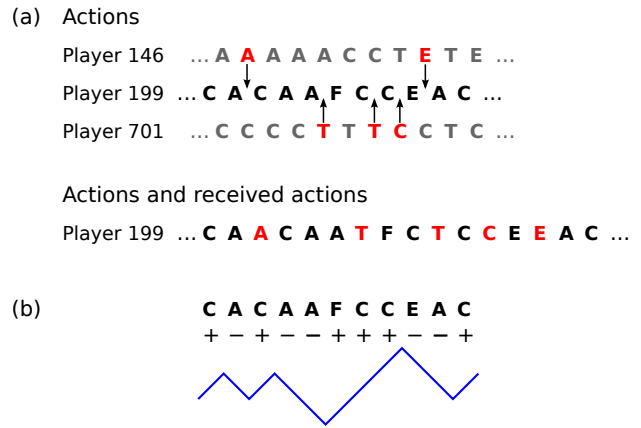}\\
    \end{center}
    \caption{(a) Short segment  of action sequences of three players, $A^{146}$, $A^{199}$, and $A^{701}$. Some actions of players 146 and 701 are directed toward player 199. 
    This results in a sequence of received-actions for player 199, $R^{199}=\{ \cdots {\rm  ATTCE } \cdots   \}$. 
    The combined sequence of actions (originated from - and directed to) player 199, $C^{199}$, is shown in the last line;  
    red letters mark actions from others directed to player 199. (b) Schematic illustration showing the definition of a binary walk in 
    `good-bad' action space (good-bad `world line'). A positive action (C, T, F or X) means an upward move, 
    a negative action (A, B, D and E) is a downward move. Good people have rising world-lines. \label{fig:seq} 
    }
   
\end{figure}


\subsection{The game}
The dataset contains practically all actions of all players of the MMOG \verb|Pardus| since  the game went online in 2004  \cite{pardus}. 
\verb|Pardus| is an open-ended online game with a worldwide player base of currently more than 370,000 people. Players live in a virtual, 
futuristic universe in which they interact with others in a multitude of ways to achieve their self-posed goals \cite{castra2005}. 
Most players engage in various economic activities typically with the (self-posed) goal to accumulate wealth and status.  
Social and economical decisions of players are often strongly influenced and driven by social factors such as 
friendship, cooperation, and conflict. Conflictual relations may result in aggressive acts such as attacks, fights, punishment, or even 
destruction of another player's means of production or transportation. 
The dataset contains longitudinal and relational data allowing for a complete and dynamical mapping of multiplex relations of the entire 
virtual society, over 1238 days. The behavioral data are free of `interviewer-bias' or laboratory effects since users are not reminded of their 
actions being logged during playing. The longitudinal aspect of the data allows for the analysis of dynamical aspects such as the emergence and evolution of 
network structures. It is possible to extract multiple social relationships between a fixed set of humans \cite{PNAS}.

\subsection{Human behavioral sequences}
  We consider eight different actions every player can execute at any time. These are  
communication (C), trade (T), setting a friendship link  (F),  removing an enemy link (forgiving) (X), 
attack (A), placing a bounty on another player (punishment) (B),  removing a friendship link (D), and setting an enemy link (E).
While C, T, F and X can be associated with {\em positive} (good) actions,   A, B, D and E are hostile or 
{\em negative} (bad) actions. We classify communication as positive because only a negligible part of communication takes place between enemies \cite{SN}.
Segments of action sequences of three players (146, 199 and 701) are shown in the first three lines of Fig.~\ref{fig:seq}~(a). 

\begin{table*}[b]
\caption*{Table 1. First row: total number of actions by all  players (with at least 1000 actions) in the Artemis universe of the Pardus game. 
Further rows: first 4 moments of $r_Y(d)$, the distribution of the log-increments of the $N_Y$ processes (see text).  Approximate 
log-normality is indicated. The large values of kurtosis for $T$ and $A$ result from a few extreme outliers. 
}
\begin{tabular}{ l|c c c c c c c c}
		& D 			& B 		&  A 		& E 		&  F 		&  C 		& T 		& X \\
		\hline
$\sum_{d=1}^{1238} N_Y(d)$	& 26,471     	&   9,914   & 558,905 &  64,444&   82,941 & 5,607,060& 393,250&  20,165\\
\hline
mean	&  	0.002   	& 0.001	& 0.004   	& -0.002	& -0.002   & 0.000    	& 0.003    & 0.002 \\
std		&    	1.13    	& 0.79   	& 0.54    	& 0.64    	& 0.35    	& 0.12   	& 0.28    	& 0.94 \\
skew		&	 0.12   	& 0.26   	& 0.35   	 & 0.08   	& 0.23    	& 0.11   	& 1.00   	& -0.01\\
kurtosis		&      3.35    	& 3.84   	& 6.23   	 & 3.67   	& 3.41    	& 3.76  	& 13.89    	& 3.72\\
\end{tabular}
\label{tab1}
\end{table*}

We consider three types of sequences  for any particular player. 
The first  is the stream of $N$ consecutive actions $A^i=\{a_n | n=1, \cdots, N \}$ 
which player $i$ performs during his `life' in the game.
The second sequence is the (time-ordered) stream of actions that player $i$ receives from  all the other players in the game, i.e. all the actions 
which are directed towards player $i$.  We denote by  $R^i=\{r_n | n=1, \cdots, M \}$  received-action sequences.
Finally, the third sequence is the time-ordered combination of player $i$'s actions and received-actions, which is 
a chronological sequence from the elements of $A^i$ and $R^i$ in the order of occurrence. The combined sequence we denote by $C^i$; its length is $M+N$, see 
also Fig.~\ref{fig:seq}~(a).  The $n$th element of one of these  series is denoted by $A^i(n)$, $R^i(n)$, or $C^i(n)$.
We do not consider the actual time between two consecutive actions which can range from milliseconds to weeks, 
rather we work in `action-time'. 

If we assign $+1$ to any positive action C, T, F or X, and $-1$ to the negative actions A, B, D and E, we can translate a
sequence $A^i$ into a symbolic binary sequence $A_{\rm bin}^i $. From the  cumulative sum of this 
binary sequence a `world line' or `random walk' for player $i$ can be generated,  
$W_{\rm good-bad}^i(t) = \sum_{n=1}^{t} A^i_{\rm bin} (n)$, see Fig.~\ref{fig:seq}~(b).  
Similarly, we define a binary sequence from the combined sequence $C^i$, where we assign $+1$  to an executed action and $-1$ to a received-action. 
This sequence we call $C^i_{\rm bin}$, its cumulative sum, $W_{\rm act-rec}^i(t) = \sum_{n=1}^{t} C^i_{\rm bin} (n)$ is the `action-receive' random-walk or world line. 
Finally, we denote the number of actions which occurred during a day in the game by $N_Y(d)$, where $d$ indicates the day and $Y$ stands for one of the eight actions.

\section{Results}
\begin{figure}[t]
    \begin{center}
            \includegraphics{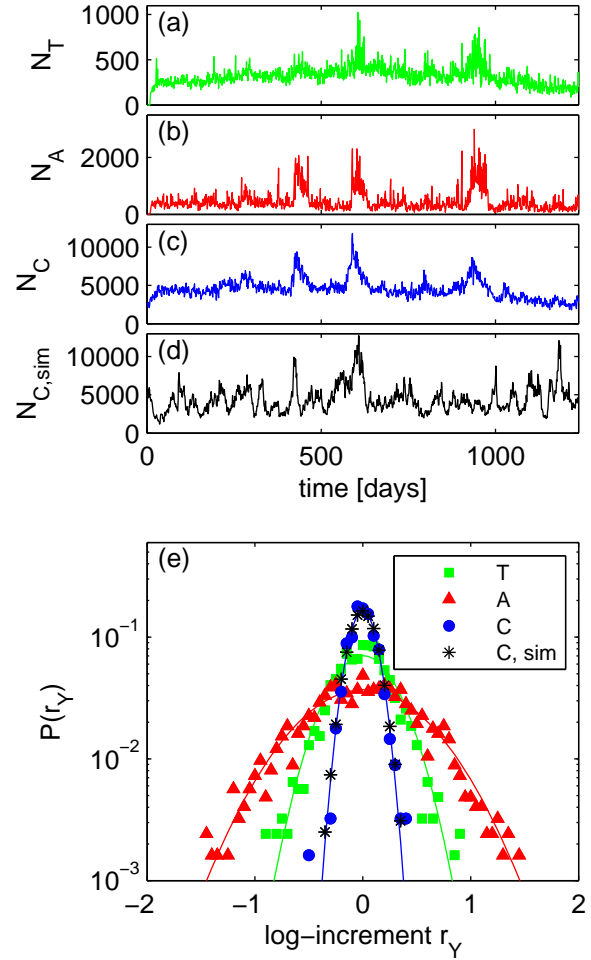}
    \end{center}
    \caption{Timeseries of the daily number of (a) trades, (b) attacks, (c) communications in the first 1238 days in the game. 
    Clearly a  mean reverting tendency of three processes can be seen. (d) Simulation of a model timeseries, Eq. (\ref{mod}), with $\rho=0.94$. We use the values  from the $N_C$
    timeseries, $R=4000$, and standard deviation $\sigma=0.12$. Compare with the  actual $N_C$ in (c). The only free parameter in the model is $\rho$. Parameters are from Tab. \ref{tab1}. Mean reversion and log-normality 
    motivate the model presented in Eq. (\ref{mod}). 
    (e) The distributions of log-increments $r_Y$ of the processes and the model. All follow approximate Gaussian distribution functions.
    \label{fig:actions}
    }
    
\end{figure}

The number of occurrences of the  various actions of all players over the entire time period is summarized in Tab. \ref{tab1} (first line).
Communication is the most dominant action, followed by attacks and trading which are each about an order of magnitude less frequent. 
The daily number of all communications, trades and attacks, $N_C(d)$, $N_T(d)$ and $N_A(d)$ is shown in Fig.  \ref{fig:actions} (a), (b) and (c), respectively. 
These processes are reverting around a mean, $R_Y$.
All processes of actions show an approximate Gaussian statistic of its log-increments, $r_Y(d)=\log \frac{N_Y(d)}{N_Y(d-1)}$.  
The first 4 moments of the $r_Y$ series are listed in Tab. \ref{tab1}.
The relatively large kurtosis for $T$ and $A$ results from a few extreme outliers. 
The distribution of log-increments for the $N_C$, $N_T$ and $N_A$ timeseries are shown in  Fig.  \ref{fig:actions} (d). 
The lines are Gaussians for the respective mean and standard deviation from Tab. \ref{tab1}.
As maybe the simplest mean-reverting model with approximate lognormal distributions, we propose 
\begin{equation}
N_Y(d)= N_Y(d-1)^{\rho_Y} \, e^{\xi(d)} \, R_Y^{(1-\rho_Y)} \quad,  
\label{mod}
\end{equation}
where $\rho_Y$ is the mean reversion coefficient, $ \xi(d)$ is a realization of a zero mean
Gaussian random number with standard deviation $\sigma_Y$, and 
$R_Y$ is the value to which the process $N_Y(t)$ reverts to. $\sigma$ is given by the third line in Tab. \ref{tab1}.

\subsection{Transition probabilities}
\begin{figure}[t]
    \begin{center}
             \includegraphics{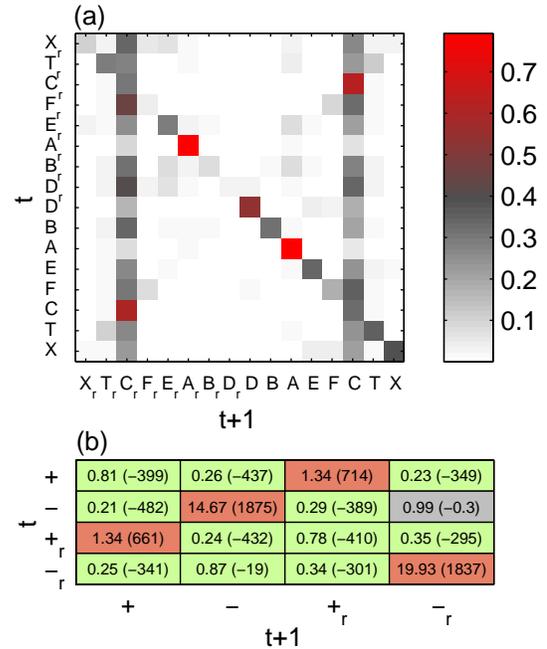}
    \end{center}
    \caption{(a) Transition probabilities $p\left( Y | Z \right)$ for actions (and received-actions) $Y$ at a time $t+1$, 
    given that a specific action $Z$ was executed or received in the previous time-step $t$.
    Received-actions are indicated by a subscript $r$.  Normalization is such that rows add up to one.
   The large values in the diagonal signal  that human actions are highly clustered or repetitive. 
   Large values for $C \rightarrow C_r$ and $C_r \rightarrow C$ reveal that communication is a tendentially anti-persistent activity -- it is more likely to receive a message after one sent a message 
   and vice versa, than to send or to receive two consecutive messages. 
   (b) The ratio $\frac{p\left( Y | Z \right)}{p(Y)}$, shows the influence of an action $Z$ at a previous time-step $t$ on a following action $Y$ at a time $t+1$, 
   where $Y$ and $Z$ can be positive or negative actions, executed or received (received actions are indicated by the subscript $r$). 
   In brackets, we report the \emph{Z}-score (significance in number of standard deviations) in respect to a sample of 100 randomized 
   versions of the dataset. The cases for which the transition probability is significantly higher (lower) than expected in uncorrelated 
   sequences are highlighted in red (green). Receiving a positive action after performing a positive action is highly overrepresented, 
   and vice versa. Performing (receiving) a negative action after performing (receiving) another negative one is also highly overrepresented. 
   Performing a negative action has no influence on receiving a negative action next. All other combinations are strongly underrepresented, 
   for example after performing a negative action it is very unlikely to perform a positive action with respect to the uncorrelated case. \label{fig:trans}
   	}
\end{figure}

With $p( Y | Z )$ we denote the probability that an action of type $Y$ follows an action of type $Z$ in the behavioral sequence of a player. $Y$ and $Z$  stand for any of the eight actions, executed or received (received is indicated by a subscript $r$). In Fig.~\ref{fig:trans}~(a) the transition probability matrix $p\left( Y|Z \right)$ is shown. The $y$ axis of the matrix indicates the action (or received-action) happening at a time $t$, the probabilities for the actions (or received-actions) that immediately follow are given in the corresponding horizontal place. 

This transition matrix specifies to which extent an action or a received action of a player is influenced by the action that was 
done or received at the previous time-step. In fact, if the behavioral sequences of players had no correlations, i.e.  
the probability of an action, received or executed, is independent of the history of the player's actions, the transition 
probability $p\left( Y | Z \right)$  simply is  $p \left( Y \right)$, i.e. to the probability that an action or received action $Y$ 
occurs in the sequence is determined by its relative frequency only. 
Therefore, deviations of the ratio $\frac{p\left( Y | Z \right)}{p(Y)}$ from $1$ indicate correlations in sequences.
In Fig.~\ref{fig:trans}~(b) we report the values of $\frac{p\left( Y | Z \right)}{p(Y)}$ for actions and received actions 
(received actions are indicated with the subscript $r$) classified only according to their positive (+) or negative (-) connotation. 
In brackets we report the \emph{Z}-score with respect to the uncorrelated case. We find that the probability to perform a good action 
is significantly higher if at the previous time-step a positive action has been received. Similarly, it is more likely that a player is the 
target of a positive action if at the previous time-step he executed a positive action. Conversely, it is highly unlikely that after a good action, 
executed or received, a player acts negatively or is the target of a negative action. Instead, in the case a player acts negatively, it is 
most likely that he will perform another negative action at the following time-step, while it is highly improbable that the following action, 
executed or received, will be positive. Finally, in the case a negative action is received, it is likely that another negative action 
will be received at the following time-step, while all other possible actions and received actions are underrepresented. 
The high statistical significance of the cases $P(-|-)$ and $P(-_r|-_r)$ hints toward a high persistence of negative actions 
in the players' behavior, see below.

Another finding is obtained by considering only pairs of received actions followed by performed actions. 
This approach allows  to quantify the influence of received actions on the performed actions of players. 
For these pairs we measure a probability of 
$0.02$
of performing a negative action after a received positive action. 
This value is significantly lower compared to the probability of $0.10$ obtained for randomly reshuffled sequences. 
Similarly, we measure a probability of $0.27$ of performing a negative action after a received negative action.  
Note that this result is not in contrast with the values in Fig.~\ref{fig:trans}~(b), since only pairs made up of received actions and performed actions are taken into account.

\begin{figure*}[t]
    \begin{center}
            \includegraphics{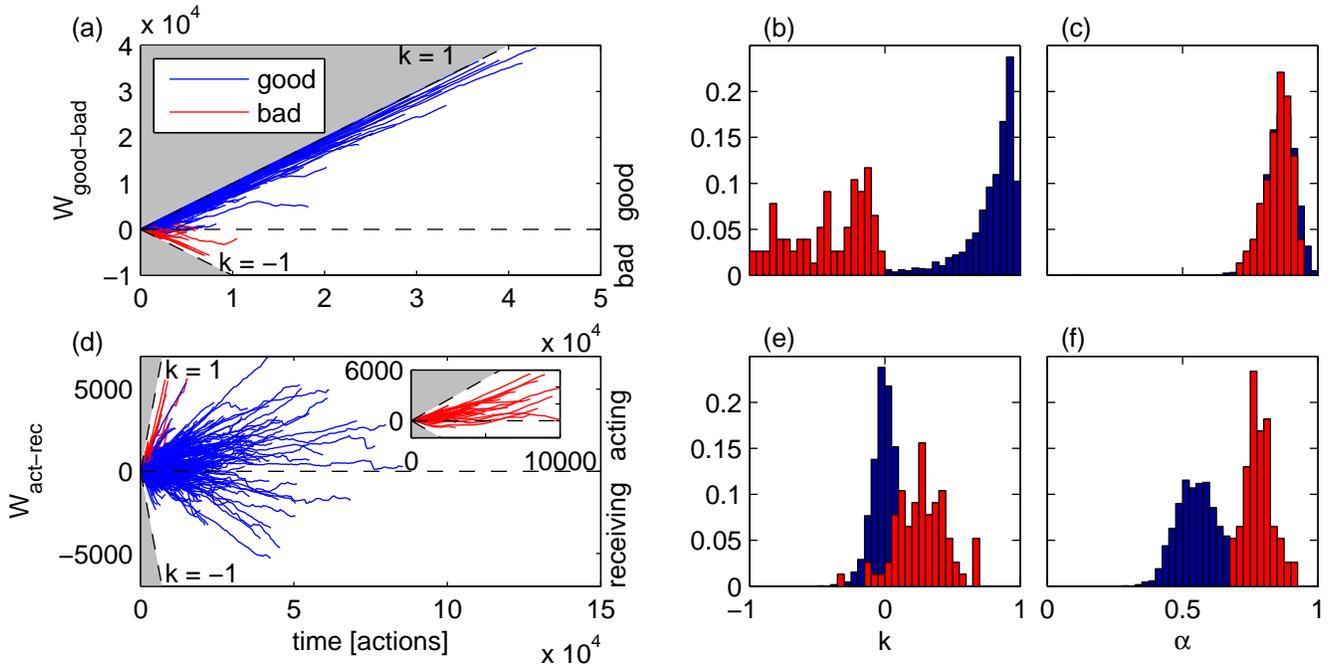}
    \end{center}
    \caption{
    (a) World lines of good-bad action random walks of the 1,758 most active players, (b) distribution of their slopes $k$ and (c) of their scaling exponents $\alpha$. By definition, players who perform more good (bad) than bad (good) actions have the endpoints of their world lines above (below) 0 in (a) and only fall into the $k>0$ ($k<0$) category in (b). (d) World lines of action-received random walks, (e) distribution of their slopes $k$ and (f) of their scaling exponents $\alpha$. The inset in (d) shows only the world lines of bad players. These players are typically dominant, i.e. they perform significantly more actions than they receive. In total the players perform many more good than bad actions and are strongly persistent with good as well as with bad behavior, see (c), i.e. actions of the same type are likely to be repeated.  \label{fig:world} }
\end{figure*}

\subsection{World lines}

The world lines $W_{\rm good-bad}^i$ of good-bad action sequences  are shown in  
Fig. \ref{fig:world} (a), the action-reaction world lines in Fig.  \ref{fig:world} (b). 

As a  simple measure to characterize these world lines we define
the slope $k$ of the line connecting the origin of the world line to its end point (last action of the player). 
A slope of $k=1(-1)$ in the good-bad world lines $W_{\rm good-bad}$ indicates that the player performed only positive (negative) actions. 
The slope $k^i$ is  an approximate measure of `altruism' for  player $i$. 
The histogram of the slopes for all players is shown in Fig.~\ref{fig:world}~(b), separated into good (blue) and bad (red) players, 
i.e. players who have performed more good than bad actions and vice versa. The mean and standard deviation of slopes of good, bad, and all players are 
$\bar k^{\rm good} = 0.81 \pm 0.19$, $\bar k^{\rm bad} = -0.40 \pm 0.28$, and $\bar k^{\rm all} = 0.76 \pm 0.31$, respectively. 
Simulated random walks with the same probability $0.90$ of performing a positive action yield 
a much lower variation, $\bar k^{\rm sim} = 0.81 \pm 0.01$, pointing at  an inherent heterogeneity of human behavior.
For the combined action--received-action  world line $W_{\rm act-rec}$ the slope is a measure of 
how well a person is integrated in her social environment. If $k=1$ the person only acts 
and receives no input, she is `isolated' but dominant. If the slope is $k=-1$ the person is driven by the actions of others and does never act nor react. The 
histogram of slopes for all players is shown in  Fig. \ref{fig:world} (e). Most players are well within the $\pm45$ degree cone. 
Mean and standard deviation of slopes of good, bad, and all players are $\bar k^{\rm good} = 0.02 \pm 0.10$, $\bar k^{\rm bad} = 0.30 \pm 0.19$, 
and $\bar k^{\rm all} = 0.04 \pm 0.12$, respectively. Bad players are tendentially dominant, i.e. they perform significantly more actions than they receive.
Simulated  random walks with equal probabilities for up and down moves for a sample of the same sequence lengths, we find again 
a much narrower distribution with slope $\bar k^{\rm sim} = 0.00 \pm 0.01$.

As a second measure we use the mean square displacement of  world lines to quantify the persistence of action sequences, 
\begin{equation}
M^2(\tau) = \langle  \Delta W (\tau) -  \langle \Delta W (\tau) \rangle  \rangle^2  \propto \tau^{2\alpha} \quad, 
\end{equation}
(see Materials and Methods). 
 The histogram of exponents  $\alpha$ 
 for the good-bad random walk, separated into good (blue) and bad (red) players, is shown in  Fig.~\ref{fig:world}~(c), for 
 the action--received-action world line in (f). In the first case strongly persistent behavior is obvious, in the second there 
 is a slight tendency towards persistence. Mean and standard deviation for the good-bad world lines 
 are $\alpha_{\rm good-bad}= 0.87\pm  0.06$,  for the action-received actions   $\alpha_{\rm act-rec}=0.59 \pm 0.10$. 
Simulated sequences of random walks have -- as expected by definition -- an exponent of $\alpha_{\rm rnd} = 0.5$, again with a very small standard deviation of about $0.02$.
Figure  \ref{fig:world} (a) also  indicates that the lifetime of players who use negative actions  frequently is short. 
The average lifetime for players with a slope $k<0$  is $2528  \pm  1856$ actions, compared to 
players with a slope $k>0$ with $3909 \pm  4559$ actions. The average 
lifetime of the whole sample of players is $3849 \pm 4484$  actions. 

\subsection{Motifs, Entropy and Zipf law}

By considering all the sequences of actions $A^i$ of all possible players $i$, we have an 
ensemble which allows to perform a motif analysis \cite{sinatra_motifs}. We define a $n$-string 
as a subsequence of $n$ contiguous actions. An $n$-motif is an $n$-string which appears in the 
sequences with a probability higher than expected, after lower-order correlations have been 
properly  removed (see Materials and Methods).

We computed the observed and expected probabilities $p^{\rm obs}$ and $p^{\rm exp}$ for all  
$8^2=64$ 2-strings and for all  $8^3=512$ 3-strings, focusing on those $n$-strings with the highest ratio $\frac{p^{\rm obs}}{p^{\rm exp}}$. 
Higher orders are statistically not feasible due to combinatorial explosion. We find that the $2$-motifs in the sequences of 
actions $A$ are clusters of same letters: BB, DD, XX, EE, FF, AA with ratios $\frac{p(\rm obs)}{p(\rm exp)} \approx 169$, $136$, $117$, $31$, $15$, $10$, respectively. 
This observation is consistent with the previous first-order observation that actions cluster. The most significant $3$-motifs however are 
(with two exceptions) the palindromes: EAX, DAF, DCD, DAD, BGB, BFB, with ratios $\frac{p(\rm obs)}{p(\rm exp)} \approx 123$, $104$, $74$, $62$, $33$, $32$, respectively. 
The exceptions disappear when one considers actions executed on the same screen in the game as equivalent, i.e. setting or removing friends or enemies: 
F, D, E, X. This observation hints towards  processes where single actions of one type tend to disrupt a flow of actions of another type.

\begin{figure}[t]
    \begin{center}
            \includegraphics{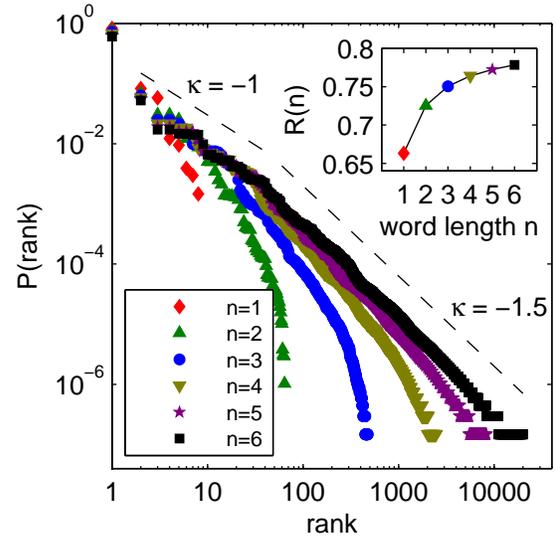}
    \end{center}
      \caption{ Rank ordered 
    probability distribution  of 1 to 6 letter words. Slopes of $\kappa = -1$ and $\kappa = -1.5$ are indicated for reference. 
    The inset shows the Shannon $n$-tuple redundancy as a function of word length $n$.
    \label{fig:entropy}
    }
    
\end{figure}

Finally, we partition the action sequences  into $n$-strings (`words'). 
Fig. \ref{fig:entropy}  shows the rank distribution of word occurrences of different lengths $n$. 
The distribution shows an approximate  Zipf law \cite{zipf} (slope of $\kappa=-1$) for ranks below 100. For ranks 
between 100 and 25,000 the scaling exponent approaches a smaller value of about $\kappa \sim -1.5$. 
The Shannon $n$-tuple redundancy (see e.g. \cite{dna, dna2, dna3}) 
for  symbol sequences composed of 8  symbols (our action types) is defined as 
\begin{equation}
 R^{(n)} = 1+ \frac{1}{3n} \sum_{i}^{8^n} P_i^{(n)} \log_2 P_i^{(n)} \quad ,
\end{equation}
where $P_i^{(n)}$ is the probability of finding a specific $n$-letter word. Uncorrelated sequences 
 yield  an equi-distribution, $P_i= 8^{-n}$, i.e. $R^{(n)} =0$. In the other extreme of only one letter being used, $R^{(n)} =1$.
In Fig. \ref{fig:entropy} (inset)  $R^{(n)}$ is shown as a function of sequence length $n$. 
Shannon redundancy is not a constant but increases with $n$.
This  indicates that Boltzmann-Gibbs entropy might not be an extensive quantity for  
action sequences \cite{hanel2011}.

\section{Discussion}

The analysis of human behavioral sequences as recorded in a massive multiplayer online game shows that communication is by far the most dominant activity
followed by aggression and trade. Communication events are about an order of magnitude more frequent than attacks and trading events, showing the importance of information exchange between humans. 
It is possible to understand the collective timeseries of human actions of a particular type ($N_Y$) with a simple mean-reverting 
log-normal model. 
On the individual level we are able to identify organizational patterns of the emergence of good overall behavior. 
Transition rates  of actions of individuals show that positive actions strongly induces  positive \emph{reactions}. 
Negative behavior on the other hand has a high tendency of being repeated instead of being reciprocated, showing the `propulsive' nature of negative actions. 
However, if we consider only reactions to negative actions, we find that negative reactions are highly overrepresented.
The probability of acting out negative actions is about 10 times higher if a person received a negative action at the previous timestep than if she received a positive action.
The action of communication is found to be of highly reciprocal `back-and-forth' nature. 
The analysis of binary timeseries of players (good-bad) shows that the behavior of almost all players is `good' almost all the time.   
Negative actions are balanced to a large extent by good ones. 
Players with a high fraction of negative actions tend to have  a significantly shorter life. 
This may be due to two reasons: First because they are hunted down by others and give up playing, second because they are unable to maintain a social life and quit the game
because of loneliness or frustration. 
We interpret these findings as empirical evidence for self organization towards reciprocal, good conduct 
within a human society. Note that the game allows bad behavior in the same way as good behavior 
but the extent of punishment of bad behavior is freely decided by the players. 

Behavior is highly persistent in terms of good and bad, as seen in the scaling exponent ($\alpha \sim 0.87$) of the mean square displacement of 
the good-bad world lines. This high persistence means that good and bad actions are carried out in clusters. Similarly high levels of persistence were 
found in a recent study of human behavior \cite{fan}. A smaller exponent ($\alpha \sim 0.59$) is found for the action--received-action timeseries. 
 
Finally we split behavioral sequences of individuals into subsequences (of length 1-6) and interpret these as behavioral `words'. 
In the ranking distribution of these words we find a Zipf law to about ranks of 100. For less frequent words the exponent in the rank distribution 
approaches a somewhat smaller exponent of about $\kappa \sim -1.5$. 
From word occurrence probabilities we further compute the Shannon $n$-tuple redundancy which yields relatively  large values when compared  for 
example  to those of  DNA sequences \cite{dna, dna2, dna3}. This reflects the dominance of communication over all the other actions. 
The $n$-tuple redundancy is clearly not a constant, reflecting again the non-trivial statistical structure of behavioral sequences. 
 
\section{Materials and Methods}
The game \verb|Pardus| \cite{pardus} is sectioned into three independent `universes'. 
Here we focus on the `Artemis' universe, in which we recorded player actions  over the first 
1,238 consecutive days of the universe's existence. Communication between any two players can 
take place directly, by using a one-to-one, e-mail-like private messaging system, or indirectly, by meeting in built-in chat channels 
or online forums. For the player action sequences analyzed we focus on one-to-one interactions  between players only, 
and discard indirect interactions such as e.g.  participation in chats or forums. 
Players can express their sympathy (distrust) toward other players by establishing so-called friendship (enmity) links. These links are 
only seen by the player marking another as a friend (enemy) and the respective recipient of that link. 
For more details on the game, see \cite{SN,pardus}.
From all sequences of all 34,055 Artemis players who performed or received an action at least once within 1,238 days, 
we removed players with a life history of less than 1000 actions, leading to the set of the most active 1,758 players which are considered throughout this work.

All data used in this study is fully anonymized; the authors have the written consent to publish from the legal department of the Medical 
University of Vienna.

\subsection{Mean square displacement}
The mean square displacement $M^2$ of a world line $W$ is defined as $M^2(\tau) = \langle  \Delta W (\tau) -  \langle \Delta W (\tau) \rangle  \rangle^2, $
where $\Delta W(\tau) \equiv W(t +\tau)- W(t)$ and $\langle . \rangle$ is the average over all  $t$. 
 The asymptotic behavior of  $M(\tau)$ yields information about the `persistence' of a  world line. $M(\tau)\propto \tau^{\frac12}$  is the  
 pure diffusion case,  $M(\tau)\propto \tau^{\alpha}$ with scaling exponent $\alpha \neq \frac12$ indicates  persistence for $\alpha>\frac12$, and anti-persistence 
 for $\alpha < \frac12$. 
 Persistence means that the probability of making an up(down)  move at time $t+1$ is larger(less) than $p=1/2$, if   the 
 move at time $t$ was an up move. For calculating the exponents $\alpha$ we use a fit range of $\tau$ between 5 and 100. We checked 
 from the mean square displacements of single world lines that this fit range is indeed reasonable.

\subsection{Motifs}
We define $n$-strings a subsequence of $n$ contiguous actions. Across the entire ensemble, $8^n$ different $n$-strings can 
appear, each of them occurring with a different probability. The frequency, or observed probability, of each $n$-string can be 
compared to its expected probability of occurrence, which can be estimated on the basis of the observed probability of lower 
order strings, i.e. on the frequency of $(n-1)$-strings. 
For example, the expected probability of occurrence of a 2-string $(A_t, A_{t+1})$ 
is estimated as the product of the observed probability of the single actions $A_t$ and 
$A_{t+1}$, $p^{\rm exp}(A_{t},A_{t+1})=p^{\rm obs}(A_{t})p^{\rm obs}(A_{t+1})$. Similarly, the probability of a 3-string 
$(A_{t}, A_{t+1}, A_{t+2})$ to occur can be estimated as $p^{\rm exp}(A_{t}, A_{t+1}, A_{t+2})=p^{\rm obs}(A_{t}, A_{t+1})p^{\rm obs}(A_{t+2}|A_{t+1})$, where $p^{\rm obs}(A_{t+2}|A_{t+1})$ is the conditional probability to have action $A_{t+2}$ following action $A_{t+1}$. By definition of conditional probability, one has $p^{\rm obs}(A_{t+2}|A_{t+1})=\frac{p^{\rm obs}(A_{t+1},A_{t+2})}{p^{\rm obs}(A_{t+1})}$ (see \cite{sinatra_motifs} for details). A $n$-motif in the ensemble is then defined as a $n$-string whose observed probability of occurrence is significantly higher than its expected probability.  
 
\begin{acknowledgments}
We thank Werner Bayer for compiling \verb|Pardus| data. Supported in part by the Austrian Science Fund, FWF P23378, and the European Cooperation in Science and Technology Action, COST MP0801.
\end{acknowledgments}

\end{article}

\end{document}